\documentclass[twocolumn,aps]{revtex4}

\newcommand {\bea}{\begin{eqnarray}}
\newcommand {\eea}{\end{eqnarray}}
\newcommand {\be}{\begin{equation}}
\newcommand {\ee}{\end{equation}}

\usepackage{graphicx}

\begin{document}


\title{The Shear Viscosity to Entropy Density Ratio\\[0.1cm] of Trapped 
Fermions in the Unitarity Limit}

\author{T.~Sch\"afer}

\affiliation{Department of Physics, North Carolina State University,
Raleigh, NC 27695}

\begin{abstract}
We extract the shear viscosity to entropy density ratio $\eta/s$ 
of cold fermionic atoms in the unitarity limit from experimental 
data on the damping of collective excitations. We find that near 
the critical temperature $\eta/s$ is roughly equal to 1/2 in units 
of $\hbar/k_B$. With the possible exception of the quark gluon plasma, 
this value is closer to the conjectured lower bound $1/(4\pi)$ than 
any other known liquid. 
\end{abstract}

\maketitle
\section{Introduction}
\label{sec_intro}

 Strongly correlated quantum systems play an important role in 
many different areas of physics. Systems of interest in atomic,
condensed matter, and nuclear physics span many orders of magnitude
in energy scale but exhibit a number of universal properties. 
Recently there has been renewed interest in the transport properties
of strongly correlated systems. Experiments at the Relativistic Heavy 
Ion Collider (RHIC) indicate that at temperatures close to the critical 
temperature $T_c$ the quark gluon plasma is strongly interacting. The 
strongly interacting quark gluon plasma (sQGP) is characterized by a 
very small shear viscosity to entropy ratio, a small heavy quark 
diffusion coefficient, and a large opacity for high energy jets 
\cite{rhic:2005,Teaney:2003kp,Shuryak:2003xe,Gyulassy:2004zy,Heinz:2005zg}.

 From a theoretical point of view not much is known about transport
coefficients of strongly correlated systems. If the interaction 
is weak then the mean free path and the shear viscosity are large.
As the strength of the interaction increases the mean free path
and the viscosity drop but there are good reasons to believe that 
the shear viscosity always remains finite. Kovtun et al.~conjectured
that there is a universal lower bound $\eta/s\geq \hbar/(4\pi k_B)$ 
\cite{Kovtun:2004de}. Here, $\eta$ is the shear viscosity, $s$ is
the entropy density, $\hbar$ is Planck's constant and $k_B$ is 
the Boltzmann constant. The bound is saturated in the case of strongly
coupled gauge theories that have a dual description in terms of 
a gravitational theory. 

 In this work we test the viscosity bound conjecture by extracting 
$\eta/s$ from experimental data on the damping of collective oscillations 
of a cold atomic gas near a Feshbach resonance. Cold atomic gases provide 
an ideal system to test the conjecture because both the temperature and
the interaction can be continuously adjusted. Also, because of universality, 
atoms in the unitarity limit are equivalent to other Fermi liquids 
with a large scattering length like dilute neutron matter. Collective 
modes in the atomic system have been studied in a number of 
experiments \cite{Kinast:2004,Kinast:2005,Bartenstein:2004,Altmeyer:2006}. 
In the weak coupling regime the frequency and damping constant of collective 
modes can be understood in terms of the Boltzmann equation. In the 
unitarity limit the frequency of collective modes is well described by 
ideal hydrodynamics \cite{Heiselberg:2004,Stringari:2004,Bulgac:2004}.
In the present work we include viscous corrections and use experimental
and quantum Monte Carlo data on the thermodynamics to extract $\eta/s$.
Previous studies of damping near the unitarity limit can be found 
in \cite{Massignan:2004,Gelman:2004,Bruun:2006}.

\section{Euler Hydrodynamics}
\label{sec_euler}

 We shall assume that the system is approximately described by ideal
(Eulerian) fluid dynamics. The equation of continuity and of momentum
conservation are given by 
\bea
\frac{\partial n}{\partial t} + \vec{\nabla}\cdot\left(n\vec{v}\right) 
 &=& 0 , \\
mn \frac{\partial \vec{v}}{\partial t} 
 + mn \left(\vec{v}\cdot\vec{\nabla} \right)\vec{v} &=& 
 -\vec{\nabla}P-n\vec{\nabla}V,
\eea 
where $n$ is the number density, $m$ is the mass of the atoms, $\vec{v}$ 
is the fluid velocity, $P$ is the pressure and $V$ is the external
potential. The trapping potential is approximately harmonic
\be 
V = \frac{m}{2}\sum_i \omega_i^2 r_i^2 .
\ee
In the unitarity limit the equation of state at zero temperature is a 
simple polytrope $P \sim n^{\gamma+1}$ with $\gamma=2/3$. At finite
temperature the equation of state is more complicated, but universality
implies that the isentropic compressibility is unaffected, 
\be
\left(\frac{\partial P}{\partial n}\right)_{S} = 
 (\gamma+1)\frac{P}{n} \ .
\ee
The equilibrium distribution $n_0$ can be determined from the hydrostatic 
equation $\vec{\nabla}P_0=-n_0\vec{\nabla}V$. At $T=0$ 
\be 
\label{ThomasFermi}
n_0(\vec{r}\,) = n_0(0) \left( 1-\sum_i \frac{r_i^2}{R_i^2}
 \right)^{1/\gamma}, \hspace{0.5cm}
 R_i^2 = \frac{2\mu}{m\omega_i^2} ,
\ee
where $\mu$ is the chemical potential. In the unitarity 
limit the chemical potential is related to the Fermi energy as $\mu=\xi 
E_F$, where $\xi$ is a universal parameter ($\xi\simeq 0.44$  according 
to the quantum Monte Carlo calculation \cite{Chang:2003}). The central 
density and the total number of particles are 
\be
\label{n0}
 n_0(0) = \frac{1}{3\pi^2} \left(\frac{2m\mu}{\xi}\right)^{3/2}, 
 \hspace{0.25cm}
 N = \frac{1}{3\xi^{3/2}}\left(\frac{\mu}{\bar\omega}\right)^3,
\ee
where $\bar\omega=(\omega_1\omega_2\omega_3)^{1/3}$. Consider small 
oscillations $n=n_0+\delta n$. From the linearized continuity and Euler 
equation we get \cite{Heiselberg:2004}
\be 
m\frac{\partial^2\vec{v}}{\partial t^2} = 
 -\gamma\left(\vec{\nabla}\cdot\vec{v}\right)
        \left(\vec\nabla V\right)
 -\vec{\nabla}\left(\vec{v}\cdot\vec{\nabla} V\right),
\ee
where we have dropped terms of the form $\nabla_i\nabla_j\vec{v}$
that involve higher derivatives of the velocity. Using a scaling 
ansatz $v_i=a_ix_i \exp(i\omega t)$ (no sum over $i$) we get 
\be 
\left(2\omega_j^2-\omega^2\right) a_j  + \gamma\omega_j^2\sum_k a_k = 0  .
\ee
This is a simple linear equation of the form $M a=0$. Non-trivial 
solutions correspond to $\det(M)=0$. In the case of a trapping 
potential with axial symmetry, $\omega_1=\omega_2=\omega_0$, 
$\omega_3=\lambda\omega_0$, we get $\omega^2 = 2\omega_0^2$ and
\cite{Heiselberg:2004,Stringari:2004,Bulgac:2004}
\bea 
\label{w_rad}
 \omega^2 &=& \omega_0^2\left\{ \gamma+1+\frac{\gamma+2}{2}\lambda^2 
  \right. \\
 && \left. \pm \sqrt{ \frac{(\gamma+2)^2}{4}\lambda^4
           + (\gamma^2-3\gamma-2)\lambda^2 
           + (\gamma+1)^2 }\right\}. \nonumber 
\eea
In the unitarity limit ($\gamma=2/3$) and for a very asymmetric trap, 
$\lambda\to 0$, the eigenfrequencies are $\omega^2=2\omega_0^2$ and 
$\omega^2=(10/3)\omega_0^2$. The mode $\omega^2=(10/3)\omega_0^2$ is 
a radial breathing mode with $\vec{a} = (a,a,0)$ and the mode 
$\omega^2=2\omega_0^2$ corresponds to a radial quadrupole $\vec{a} = 
(a,-a,0)$.

\begin{figure}
\begin{center}
\hspace{-0.5cm}
\includegraphics[width=8.5cm]{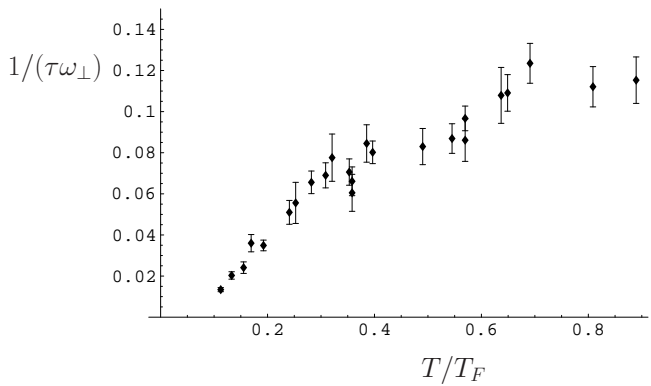}
\end{center}
\caption{\label{fig_damp}
Temperature dependence of the damping rate for the radial breathing 
mode of a trapped $^6$Li gas near the Feshbach resonance at 840 G, 
from Kinast et al.~\cite{Kinast:2005}. Here $\tau$ is the damping time 
and $\omega_\perp$ is the radial trap frequency. We have used the
calibration curve in \cite{Kinast:2005b} to convert the $\tilde{T}$ scale 
of \cite{Kinast:2005} to $T/T_F$.   }
\end{figure}

\section{Viscous Corrections}
\label{sec_visc}

 The energy dissipated due to viscous effects is 
\bea 
\dot{E} &=& -\frac{1}{2} \int d^3x\, \eta(x)\, 
  \left(\partial_iv_j+\partial_jv_i-\frac{2}{3}\delta_{ij}
      \partial_k v_k \right)^2  \\
 & & \mbox{}
   -\int d^3x \, \zeta(x)\, \big( \partial_iv_i\big)^2 
  \nonumber , 
\eea
where $\eta$ is the shear viscosity and $\zeta$ is the bulk viscosity.
In the unitarity limit the system is scale invariant and the bulk 
viscosity in the normal phase vanishes. The situation is more complicated 
in the superfluid phase. In this case the normal and superfluid components 
can flow independently and in addition to the shear viscosity there are 
three bulk viscosities $\zeta_i$. Two of the three bulk viscosities, 
$\zeta_1$ and $\zeta_2$, can be shown to vanish as a consequence of
scale invariance \cite{Son:2005}. The third bulk viscosity, $\zeta_3$, 
only contributes to dissipation if the normal components is moving 
relative to the superfluid \cite{Khalatnikov}. In the following we 
shall consider modes in which the two components move in phase and 
ignore bulk viscosity. We will also assume that the viscosity only 
depends on $x$ through the local density and temperature. This is 
valid if the density and temperature are varying slowly, and is 
consistent with the local density approximation for the density 
profile. We note that in this approximation the equation of state
and the transport coefficients reflect the conformal invariance
of the microscopic dynamics, even though scale invariance is
broken by the external potential. 

For the radial scaling flows given in equ.~(\ref{w_rad}) we have 
\be 
\overline{\dot{E}} = -\frac{2}{3}
  \left( a_x^2+a_y^2-a_xa_y \right) \int d^3x\, \eta(x),
\ee
where $\overline{E}$ is a time average. The damping rate 
is determined by the ratio of the energy dissipated to the total
energy of the collective mode. The kinetic energy is 
\be 
 E_{kin} = \frac{m}{2}\, \int d^3x\, n(x) \vec{v}^2 
 = \frac{mN}{2} \left( a_x^2+a_y^2 \right) 
  \langle x^2 \rangle. 
\ee
At $T=0$ we find $\langle x^2 \rangle = R_\perp^2/8$, where $R_\perp$ 
is the transverse size of the cloud. At non-zero temperature we 
can use the Virial theorem \cite{Thomas:2005} to relate $\langle x^2 
\rangle$ to the total energy of the equilibrium state, $\langle x^2
\rangle/\langle x^2\rangle_{T=0}=E/E_{T=0}$. The damping rate is 
\cite{Kavoulakis:1998,Gelman:2004}
\be 
\label{edot_e}
-\frac{1}{2} \frac{\overline{\dot{E}}}{E}   = 
 \frac{2}{3}\, \frac{a_x^2+a_y^2-a_xa_y}{a_x^2+a_y^2}\,
 \frac{\int d^3x\, \eta(x)}{mN \langle x^2\rangle} . 
\ee
Note that the second factor on the RHS is $1/2$ for the radial breathing 
mode and $3/2$ for the radial quadrupole mode. If this dependence could be 
demonstrated experimentally, it would confirm that the damping is indeed 
dominated by shear stress. Another possibility is to compare the breathing 
mode with a scissors mode. The scissors mode is characterized by the 
velocity field $\vec{v}=a\vec{\nabla}(xy)$. The frequency is $\omega^2=
\omega_x^2+\omega_y^2$ and the second factor in equ.~(\ref{edot_e}) is 6. 

 For the unitary Fermi gas the ratio of the shear viscosity to the entropy 
density is given by universal function that depends only on the ratio $T/T_F$, 
$\eta(\mu,T)=\alpha(T/T_F)s(\mu,T)$. Here, the Fermi temperature is given 
by $T_F=(3\pi^2n)^{2/3}/(2m)$. In the local density approximation this 
implies that $\eta(x)=\alpha (T/T_F(x))s(x)$, where $T_F(x)$ is the local 
Fermi temperature. We shall assume that $\alpha$ is a smooth function and
replace $T_F(x)$ by its value at the center of the trap. This approximation
can be checked a posteriori. We note that since the flow profile has a 
simple scaling form the damping rate is proportional to the volume integral 
of the shear viscosity. If $\eta\sim s$  then the damping rate scales 
with the total entropy. The kinetic energy, on the other hand, is 
proportional to the number of particles. The shear viscosity to entropy 
density ratio extracted from the radial breathing mode is 
\be 
\label{eta_s}
\frac{\eta}{s}  =\frac{3}{4} \xi^{1/2} (3N)^{1/3} 
 \left(\frac{\bar\omega\Gamma}{\omega_\perp^2}\right)
 \left(\frac{E}{E_{T=0}}\right)
 \left(\frac{N}{S}\right),
\ee
where $\Gamma/\omega_\perp=1/(\tau\omega_\perp)$ is the dimensionless
damping rate.

\begin{figure}
\begin{center}
\includegraphics[width=8cm]{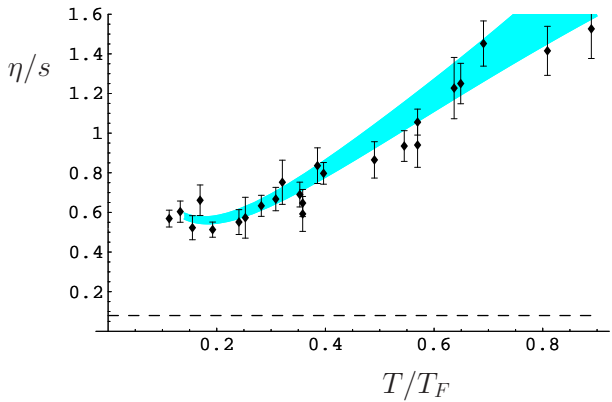}
\end{center}
\caption{\label{fig_eta_s}
Viscosity to entropy density ratio of a cold atomic gas in the 
unitarity limit. This plot is based on the damping data 
published in \cite{Kinast:2005} and the thermodynamic data in 
\cite{Kinast:2005b,Luo:2006}. The dashed line shows the 
conjectured viscosity bound $\eta/s=1/(4\pi)$. The shaded band
is a systematic error estimate based on the contribution to 
$\dot{E}$ from atoms outside a surface at optical depth one. }
\end{figure}

\section{Breakdown of Hydrodynamics}
\label{sec_br}

 Near the edge of the cloud the mean free path is comparable to the 
characteristic length scale over which the velocity field varies and 
the hydrodynamic description breaks down. In this regime the density 
is low and the mean free path can be estimated using kinetic theory. In the 
limit $n\ll (mT)^{3/2}$ the mean free path is given by 
\cite{Bruun:2006}
\be 
\label{mfp}
 l_{\it mfp}=(n\sigma)^{-1}, \hspace{1cm} 
 \sigma = \frac{4\pi}{3}\,\frac{1}{mk_BT}.
\ee
Following \cite{Kavoulakis:1998} we can define a surface $r_0(\theta)$
by the condition that a particle incident from the exterior of the 
cloud has a chance of no larger than $1/e$ of colliding with another 
particle
\be 
\label{opt_d}
 1 = \int_{r_0(\theta)}^\infty 
\frac{ds}{l_{\it mfp}(\vec{r}\,)} \; . 
\ee
Here, $\theta$ denotes the angle between $ds$ and the $z$-axis. Ideally, 
the hydrodynamic description inside the surface $r_0(\theta)$ should 
be coupled to kinetic theory outside this surface. In this work we
shall be less ambitious and use equ.~(\ref{opt_d}) in order to 
estimate the systematic uncertainty in our determination of $\eta/s$. 
For this purpose we compute the contribution to $\dot{E}$ in 
equ.~(\ref{edot_e}) that comes from atoms outside $r_0(\theta)$
and treat it as a systematic error in the damping rate. 

\section{Thermodynamics}
\label{sec_eos}

  Experimental results for $\Gamma/\omega$ are shown in 
Fig.~(\ref{fig_damp}). In order to extract $\eta/s$ we also need
information on the entropy per particle. In the unitarity limit 
there are only two energy scales in the problem, the chemical 
potential $\mu$ and the thermal energy $k_BT$ (in the following 
we will set $k_B$ to one). The associated density scales are 
$n_{f}(\mu)$ and $\lambda_{T}^{-3}$, where $n_{f}(\mu)=(2m\mu)^{3/2}
/(3\pi^{2})$ is the density of a free Fermi gas and $\lambda_{T}=
(2\pi mT)^{-1/2}$ is the thermal wave length. All thermodynamic 
quantities can be expressed as suitable powers of either $n_{f}(\mu)$ 
or $\lambda_{T}^3$ times a function of the dimensionless quantity 
$y=T/\mu$. For example, we can write the pressure as \cite{Ho:2004}
\be
\label{p_uni}
P(\mu,T)=\frac{2}{5} \mu n_{f}(\mu) \mathcal{G}(y)
        =\frac{2}{5} T\lambda_T^{-3}\mathcal{W}(y^{-1})\, , 
\ee
where the first form is more useful at small $T$ and the second
at high $T$. Using standard thermodynamic identities one can show 
that
\bea
\label{n_uni}
n(\mu,T)&=&n_{f}(\mu)\mathcal{F}(y), \\
s(\mu,T)&=&\frac{2}{5}n_{f}(\mu)\mathcal{G}^{\prime}(y),
\eea
where $n$ is the density, $s$ is the entropy density, and 
$\mathcal{F}(y)=\mathcal{G}(y)-2y\mathcal{G}^{\prime}(y)/5$.
At $T=0$ the function $\mathcal{G}(y)$ is related to the 
parameter $\xi$ introduced above, $\mathcal{G}(0)=\xi^{-3/2}$.
The functions $\mathcal{G}(y)$, $\mathcal{F}(y)$ refer to the 
bulk system. Trapped systems can be described using the local
density approximation. Experiments typically involve $10^5-10^6$
atoms and the local density approximation is very accurate. The
density of the trapped system is 
\be 
\label{lda}
 n_0(x) = n\left(\mu-V(x),T\right),
\ee
where $V(x)$ is the trapping potential. Similar relations hold
for the energy and entropy density. For $T=0$ equ.~(\ref{lda})
reduces to equ.~(\ref{ThomasFermi}).

 The function $\mathcal{G}(y)$ can be extracted from quantum 
Monte Carlo data or from calorimetric experiments with trapped 
fermions. A number of Monte Carlo calculations have appeared over 
the last couple of years \cite{Lee:2005,Burovski:2006,Bulgac:2006}, 
but there are still significant disagreements between the results. 
Burovski et al.~find $T_c/T_F=0.152(7)$ and a critical entropy per 
particle $S/N=0.16(2)$ \cite{Burovski:2006}. Bulgac et al.~quote
$T_c/T_F=0.23(2)$ and $S/N\simeq 1.1$ \cite{Bulgac:2006}. Using the
local density approximation the results of Bulgac et al.~correspond
to a critical entropy per particle of $S/N\simeq 2.1$ for the 
trapped system. The reason that the entropy per particle is larger 
in a trapped system is that the density near the edges is smaller, and 
therefore the dimensionless temperature $mT/n^{2/3}$ larger. 

 The Duke group has performed a series of calorimetric measurements
\cite{Kinast:2005b,Luo:2006}. Kinast et al.~\cite{Kinast:2005b} 
provide a simple parametrization of the energy of the trapped 
system as a function of $t=T/T_F$. The result is 
\be 
E=E_0\,\left\{
\begin{array}{ll}
1 + 97.3\,t^{3.73}\;\;  &  t<t_c , \\
1 + 4.98\,t^{1.43}\;\;  &  t>t_c ,
\end{array}\right.
\ee 
with $E_0=0.53 E_F$ and $T_c/T_F=0.29(2)$. Luo et al.~\cite{Luo:2006} 
give a similar parametrization of the entropy, 
\be 
S/N= \,\left\{
\begin{array}{ll}
4.6\,(e-e_0)^{0.61} \;\; & e<e_c ,\\
4.0\,(e-e_0)^{0.45} \;\; & e>e_c ,
\end{array}\right.
\ee
where $e=E/E_F$. The critical entropy per particle is $S/N=2.7$, 
roughly compatible with the Monte Carlo results of Bulgac et al., 
but significantly larger than the results of Burovski et al.

\section{Results and conclusions}
\label{sec_sum}

 In Fig.~\ref{fig_eta_s} we show $\eta/s$ extracted using both 
the damping data and the calorimetry from the Duke group. We 
observe that $\eta/s$ is small ($\sim 0.5$) near $T_c$ and
slowly grows with temperature for $T>T_c$. The value of $\eta/s$ 
near $T_c$ is about six times larger than the conjectured viscosity 
bound and consistent with the picture of a very strongly correlated 
liquid. Indeed, the extracted value $\eta/s\sim 0.5$ is smaller than 
the previously known minimum for all other liquids, $\eta/s\sim 0.7$
for liquid Helium near the lambda point \cite{Kovtun:2004de}. Even 
smaller values of $\eta/s$ have been reported for the quark gluon 
plasma produced at RHIC, but the uncertainties remain large 
\cite{Teaney:2003kp,Hirano:2005wx}. We note that the systematic
uncertainty near the minimum is small, but that it increases as 
a function of temperature. 

 There are a number of issues that need to be addressed in more 
detail. We argued that it is important to establish that shear 
viscosity is indeed the dominant damping mechanism. This can be 
done either by studying the dependence of the damping time on
the type of collective mode, or by studying the dependence on 
system size. Kinast et al.~collected some data on system size 
dependence below $T_c$ and find that $\Gamma/\omega_\perp$ is 
roughly independent of the number of particles \cite{Kinast:2005}. 
This is not consistent with the scaling in equ.~(\ref{eta_s}).
Data on system size dependence can also be used to determine 
at what point hydrodynamics is breaking down. 

 It is also important to understand the damping mechanism below 
$T_c$ in more detail. The data show a very simple linear behavior
in the variable $\tilde{T}\sim (T/T_F)^{2/3}$. The natural framework 
for understanding the damping mechanism in the regime below $T_c$ is 
superfluid (two-fluid) hydrodynamics \cite{Khalatnikov}. There are 
several sound modes in superfluid hydrodynamics. First sound is an 
excitation in which the superfluid and normal components move together, 
whereas second sound corresponds to an oscillation of the superfluid 
component against the normal one. The damping of first sound is 
governed by the shear stress of the normal component. It is likely 
that the collective excitations that have been observed experimentally 
are ordinary (first) sound modes, but it is not obvious why the damping 
constant is linear in $\tilde{T}$. We should note that a linear behavior 
was observed in trapped Bose gases \cite{Chevy:2002}, where it was 
attributed to Landau damping by normal excitations \cite{Fedichev:1998}.

 Finally, there are a number of technical aspects of our analysis 
that should be improved. We have assumed that the quantity 
$\alpha=\eta/s$ is only weakly temperature dependent. Near the 
the minimum of $\eta/s$ this is a good approximation, but at 
higher temperature the uncertainty inherent in this approximation 
grows, as does the uncertainty related to the breakdown of hydrodynamics 
near the the surface of the cloud. In this regime a Boltzmann description 
should be used. In the present work we have neglected dissipation due to 
temperature gradients and thermal conductivity. This is expected to be 
a good approximation for scaling flows because oscillations in density 
are proportional to the equilibrium density $\delta n(x)\sim n_0(x)$. 
For isentropic oscillations $\delta T \sim (\delta n/n)T$ and to
leading order no temperature gradients are present.
 
Acknowledgments: This work is supported in part by the US Department
of Energy grant DE-FG02-03ER41260. I would like to thank Dam Son 
for useful correspondence, and John Thomas and Andrey Turlapov
for pointing out an error in an earlier draft.


\end{document}